\documentclass[fleqn,12pt,twoside]{article}
\usepackage{espcrc1}
\usepackage{epsf,graphics,rotate}

\newcommand{\insertfig}[2]{\mbox{\epsfxsize=#1cm \epsfbox{#2.eps}}}
\newcommand{\Bx}{x_{\rm B}}

\newcommand{\GeV}{\mbox{GeV}}

\newcommand{\AmS}{{\protect\the\textfont2A\kern-.1667em\lower.5ex\hbox{M}\kern-.125emS}}

\begin{document}

\thispagestyle{empty}

\begin{flushright}
DOE/ER/40762-259 \\
UMD-PP\#02-059 \\
\end{flushright}

\vspace{5mm}

\centerline{\large \bf Nucleon hologram with exclusive leptoproduction}

\vspace{10mm}

\centerline{\bf A.V. Belitsky$^a$, D. M\"uller$^b$}

\vspace{5mm}

\centerline{\it $^a$Department of Physics}
\centerline{\it University of Maryland at College Park}
\centerline{\it College Park, MD 20742-4111, USA}

\vspace{5mm}

\centerline{\it $^a$Fachbereich Physik}
\centerline{\it Universit\"at Wuppertal}
\centerline{\it D-42097 Wuppertal, Germany}

\vspace{15mm}

\centerline{\bf Abstract}

\vspace{5mm}

Hard exclusive leptoproductions of real photons, lepton pairs and mesons are
the most promising tools to unravel the three-dimensional picture of the
nucleon, which cannot be deduced from conventional inclusive processes like
deeply inelastic scattering.

\vspace{40mm}

\centerline{\it Brief summary of talks given at}
\centerline{\it 17th Summer School HUGS@CEBAF (Jefferson Lab, June 2002)}
\centerline{\it Workshop ``Exclusive processes at high momentum transfer"
                (Jefferson Lab, May 2002)}
\centerline{\it Workshop ``QCD structure of the nucleon" (Ferrara, April 2002)}
\centerline{\it Workshop ``Testing QCD through spin observables"
                (Charlottesville, April 2002)}
\centerline{\it Conference ``Baryons 2002" (Jefferson Lab, March 2002)}
\centerline{\it ``Electron-Ion Collider Workshop" (Brookhaven Lab, February 2002)}

\setcounter{page}{0}


\title{Nucleon hologram with exclusive leptoproduction}

\author{
A.V. Belitsky\address[MCSD]{
Department of Physics, University of Maryland, College Park, MD 20742-4111, USA},
D. M\"uller\address[MCSD]{
Fachbereich Physik, Universit\"at Wuppertal, D-42097 Wuppertal, Germany}
}


\maketitle

\begin{abstract}

Hard exclusive leptoproductions of real photons, lepton pairs and mesons are
the most promising tools to unravel the three-dimensional picture of the
nucleon, which cannot be deduced from conventional inclusive processes like
deeply inelastic scattering.

\end{abstract}

\section{From macro to micro}

\begin{figure}[t]
\begin{center}
\mbox{
\begin{picture}(0,53)(225,0)
\put(0,-32){\insertfig{7.5}{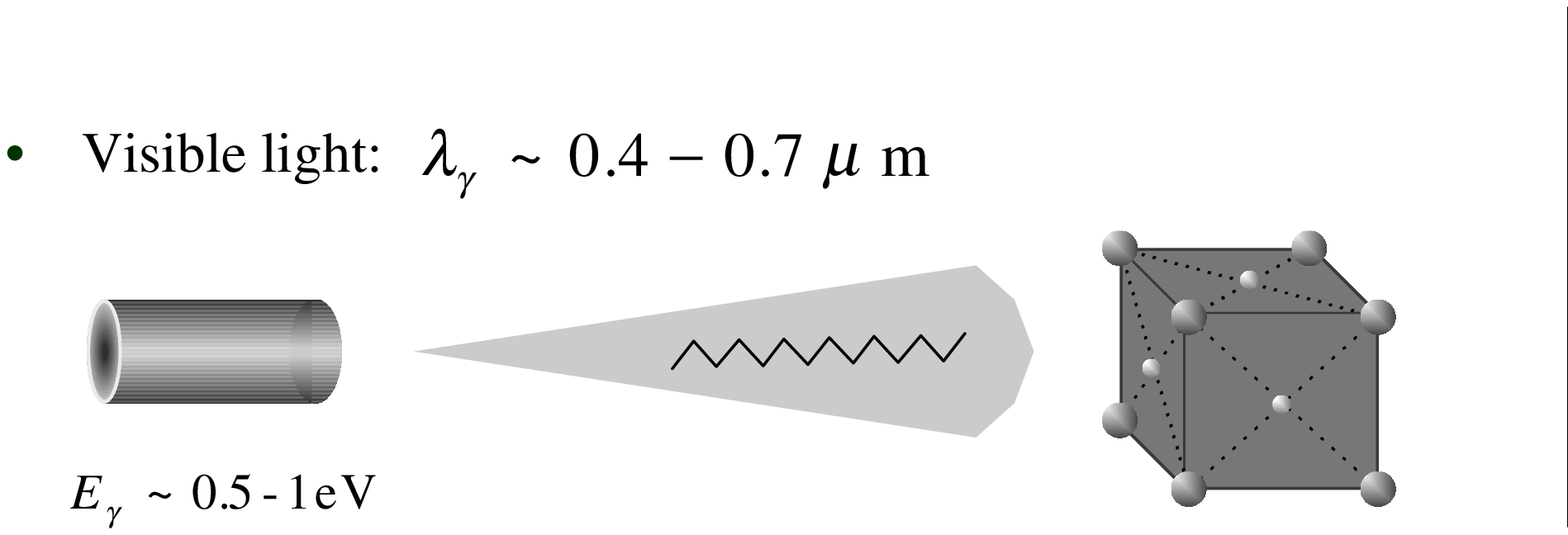}}
\put(235,-32){\insertfig{7.5}{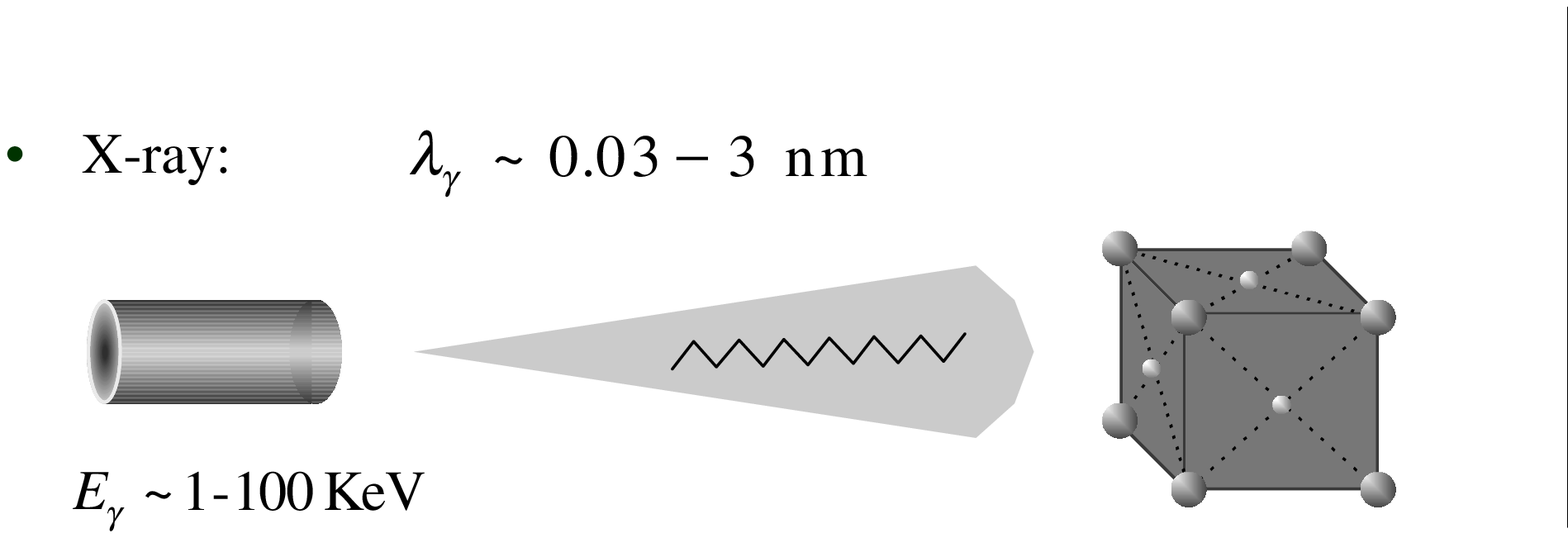}}
\end{picture}
}
\end{center}
\caption{\label{Crystalography} Left: A beam of visible light does not
resolve the crystal's structure. Right: An X-ray beam does and creates
a diffraction pattern on the photo-plate.}
\vspace{-0.5cm}
\end{figure}

Why do we see the world around us the way it is? Human eyes can detect
electromagnetic waves in a very narrow range of wavelength, $\lambda_\gamma
\sim 0.4 - 0.7 \mu{\rm m}$, which we call visible light. The light
from a source, say the sun, is reflected from the surface of macro-objects
and is absorbed by the eye's retina which transforms it into a neural
signal going to the brain which forms the picture. The same principle is
used in radars which detect reflected electromagnetic waves of a meter
wavelength. The only requirement to ``see" an object is that the length
of resolving waves must be comparable to or smaller than its size. The
same conditions have to be obeyed in case one wants to study the microworld,
e.g., the structure of macromolecules (DNA, RNA) or assemblies (viruses,
ribosomes). Obviously, when one puts a chunk of material in front of a
source of visible light, see Fig.\ \ref{Crystalography}, the object merely
leaves a shadow on a screen behind it and one does not see its elementary
building blocks, i.e., atoms. Obviously, visible light is not capable to
resolve the internal lattice structure of a crystal since the size of
an individual atom, say hydrogen, is of order $r_{\rm atom} \sim \left(
\alpha_{\rm em} m_e \right)^{-1} \sim \left( 10 \ {\rm KeV} \right)^{- 1}$
and the light does not diffract from it. Therefore, to ``see" atoms in
crystals one has to have photons with the wavelength $\lambda_\gamma \leq
r_{\rm atom}$, or equivalently, of the energy $E_\gamma \geq r^{-1}_{\rm atom}$.
To do this kind of ``nano-photography" one needs a beam of X-rays which after
passing through the crystal creates fringes on a photo-plate, see Fig.\
\ref{Crystalography}. Does one get a three-dimensional picture from such a
measurement? Unfortunately, no. In order to reconstruct atomic positions in
the crystal's lattice one has to perform an inverse Fourier transform. This
requires knowledge of both the magnitude and the phase of diffracted waves.
However, what is measured experimentally is essentially a count of number of
X-ray photons in each spot of the photo-plate. The number of photons gives
the intensity, which is the square of the amplitude of diffracted waves. There
is no practical way of measuring the relative phase angles for different
diffracted spots experimentally. Therefore, one cannot unambiguously reconstruct
the crystal's lattice. This is termed as ``The Phase Problem". None of techniques
called to tackle the problem provides a parameter-free answer.

When we study hadronic matter at the fundamental level we attempt to perform
the ``femto-photography" of the interior constituents (quarks and gluons) of
strongly interacting ``elementary" particles such as the nucleon. Quantum
$\chi\rho\omega\mu\alpha$ dynamics, the theory of strong interaction, is not
handy at present to solve the quark bound state problem. Therefore,
phenomenological approaches, based on accurate analyses of high-energy
scattering experimental data and making use of rigorous perturbative QCD
predictions, are indispensable for a meticulous understanding of the
nucleon's structure. As we discuss below most of high-energy processes
resolving the nucleon content, such as described in terms of form factors
and inclusive parton densities, suffer from the same ``Phase Problem" and
therefore they lack the opportunity to visualize its three-dimensional
structure. A panacea is found in newborn generalized parton distributions
\cite{MueRobGeyDitHor94}, which are measurable in exclusive leptoproduction
experiments.

\section{Form factors}

Nucleon form factors are measured in the elastic process $\ell N \to \ell'
N'$. Its amplitude is given by the lepton current $L_\mu (\Delta) \equiv \bar
u_\ell (k - \Delta) \gamma_\mu u_\ell (k)$ interacting via photon exchange with
the nucleon matrix element of the quark electromagnetic current $j_\mu (x) =
\sum_q e_q \bar q (x) \gamma_\mu q (x)$:
\begin{equation}
\label{FormFactor}
{\cal A}_{NN'}
=
\frac{1}{\Delta^2} L_\mu (\Delta) \langle p_2 | j_\mu (0) | p_1 \rangle
\equiv
\frac{1}{\Delta^2} L_\mu (\Delta)
\left\{
h_\mu F_1 (\Delta^2)
+
e_\mu
F_2 (\Delta^2)
\right\}
\, .
\end{equation}
Here the matrix element of the quark current is decomposed in terms
of Dirac and Pauli form factors ($\Delta \equiv p_2 - p_1$), accompanied by
the Dirac bilinears $h_\mu \equiv \bar u_N (p_2) \gamma_\mu u_N (p_1)$ and
$e_\mu \equiv \bar u_N (p_2) i \sigma_{\mu\nu} \Delta_\nu u_N (p_1)/(2 M_N)$.
In the Breit frame $\vec p_2 = - \vec p_1 = \vec \Delta / 2$ there is no
energy exchange $E_1 = E_2 = E$ and thus relativistic effects are absent.
The momentum transfer is three-dimensional $\Delta^2 = - \vec \Delta^2$, so
that
\begin{eqnarray}
\langle p_2 | j_0 (0) | p_1 \rangle
=
\tilde\varphi^\ast_2 \tilde\varphi_1 G_E (- \vec\Delta^2) \, , \quad
\langle p_2 | \vec j (0) | p_1 \rangle
=
- \frac{i}{2 M_N}
\tilde\varphi^\ast_2
[ \vec\Delta \times \vec\sigma ]
\tilde\varphi_1 G_M (- \vec\Delta^2) \, ,
\end{eqnarray}
are expressed in terms of Sachs electric $G_E (\Delta^2) \equiv F_1 (\Delta^2)
+ \Delta^2 / (4M_N^2) F_2 (\Delta^2)$ and magnetic $G_M (\Delta^2) \equiv F_1
(\Delta^2) + F_2 (\Delta^2)$ form factors. Introducing the charge $q \equiv
\frac{1}{V} \int d^3 \vec x \, j_0 (\vec x)$ and magnetic moment $\vec \mu
\equiv \frac{1}{V} \int d^3 \vec x \, [ \vec x \times \vec j ] (x)$
operators, one finds the normalization
\begin{equation}
\langle p | q | p \rangle
=
\tilde\varphi^\ast_2 \tilde\varphi_1 G_E (0) \, , \quad
\langle p | \vec\mu | p \rangle
=
\frac{\tilde\varphi^\ast_2 \vec\sigma \tilde\varphi_1}{2 M_N} G_M (0) \, .
\end{equation}

The interpretation of Sachs form factors as Fourier transforms of charge and
magnetization densities in the nucleon requires to introduce localized
nucleon states in the position space $| \vec x \rangle$ as opposed to the
plane-wave states used above $| p \rangle$,
\begin{equation}
| \vec x \rangle
=
\sum_{\vec p} \frac{{\rm e}^{i \vec p \cdot \vec x}}{\sqrt{V}}
{\mit\Psi} (\vec p) | \vec p \rangle
\, ,\quad \mbox{with}\quad  \sum_{\vec p} |{\mit\Psi} (\vec p)|^2
= 1\, .
\end{equation}
Here a very broad wave packet  ${\mit\Psi} (\vec p) \approx \, \mbox{const}$
is assumed in the momentum space. Then the charge density $\rho (\vec x)$
of the nucleon, localized at $\vec x = 0$, is
\begin{equation}
\langle \vec x = 0 | j_0 (\vec x)| \vec x = 0 \rangle
\equiv
\tilde\varphi^\ast_2 \tilde\varphi_1 \rho (\vec x)
=
\tilde\varphi^\ast_2 \tilde\varphi_1
\int \frac{d^3 \vec \Delta}{(2 \pi)^3} {\rm e}^{- i \vec\Delta \cdot \vec x}
G_E (- \vec\Delta^2) \, ,
\end{equation}
and similar for the magnetic form factor. The famous Hofstadter's experiments
established that the proton is not a point-like particle $\rho_{\rm point}
(\vec x) = \delta^3 (\vec x)$ which would have $G^{\rm point}_E = \mbox{const}$,
but rather $G_E (- \vec\Delta^2) \approx ( 1 + \vec\Delta^2 r_N^2 / 6 )^{- 2}$
with the mean square radius $r_N \approx 0.7 \, \mbox{fm}$.

\begin{figure}[t]
\begin{center}
\mbox{
\begin{picture}(0,148)(215,0)
\put(0,1){\insertfig{2.65}{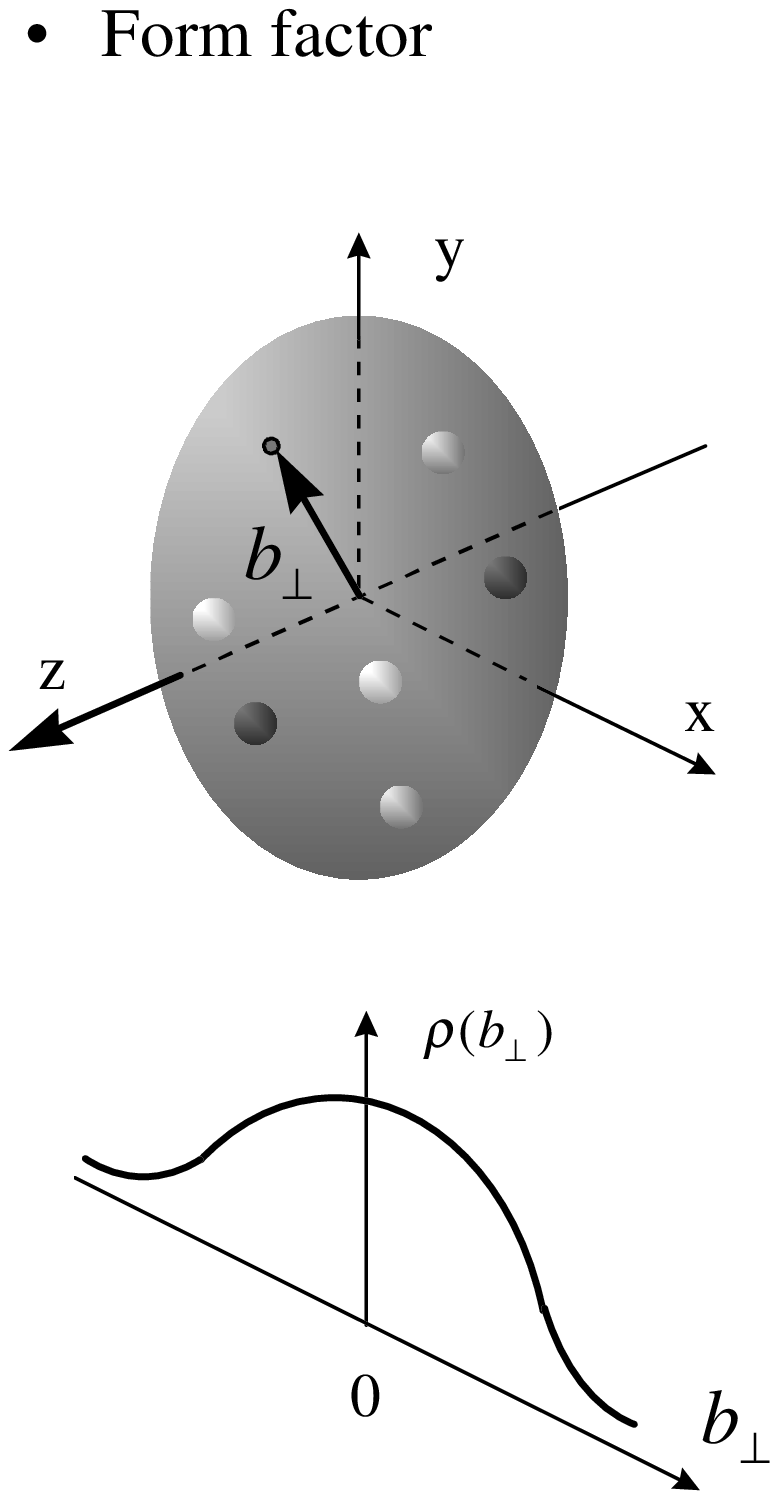}}
\put(160,-16){\insertfig{2.9}{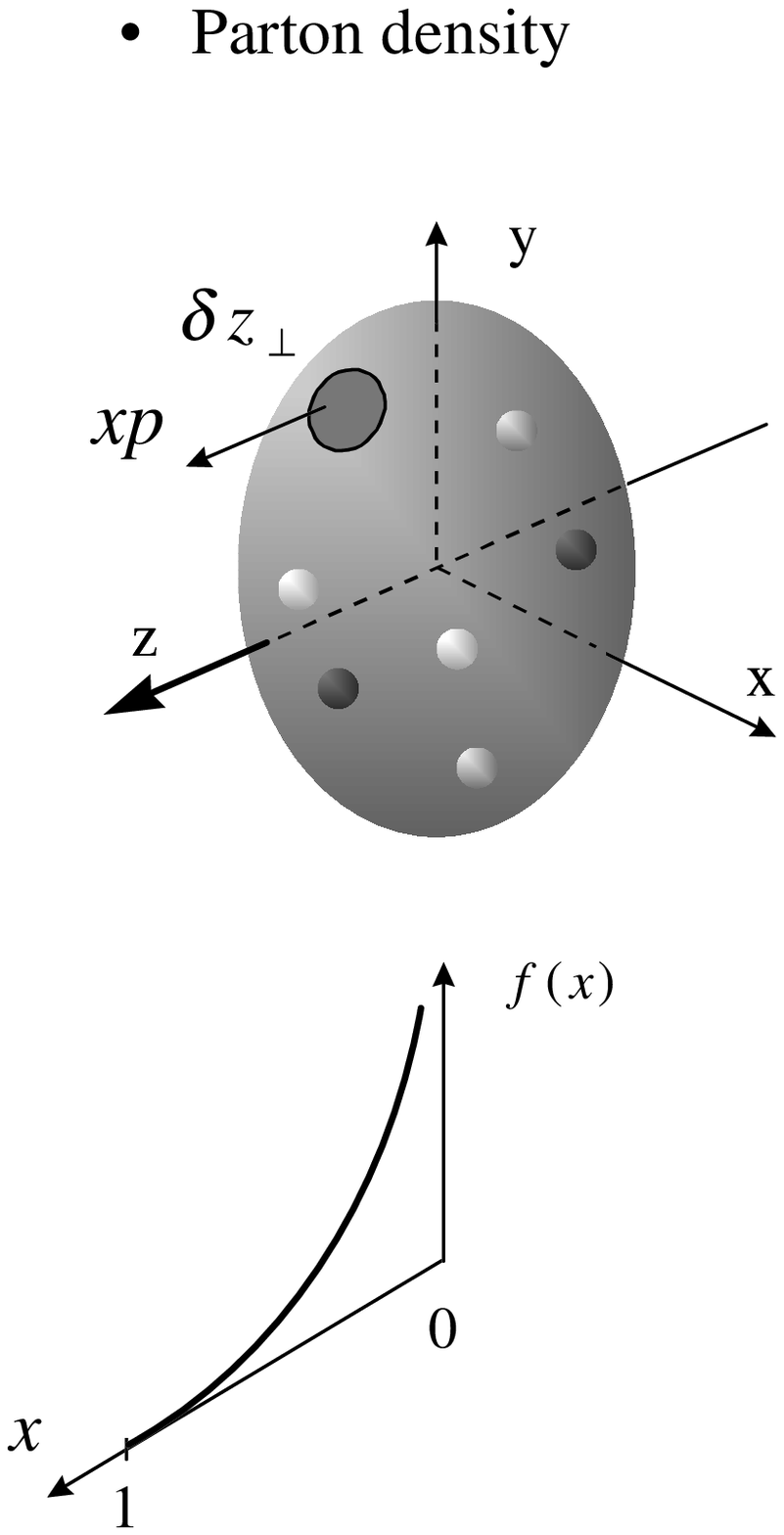}}
\put(320,-34){\insertfig{4}{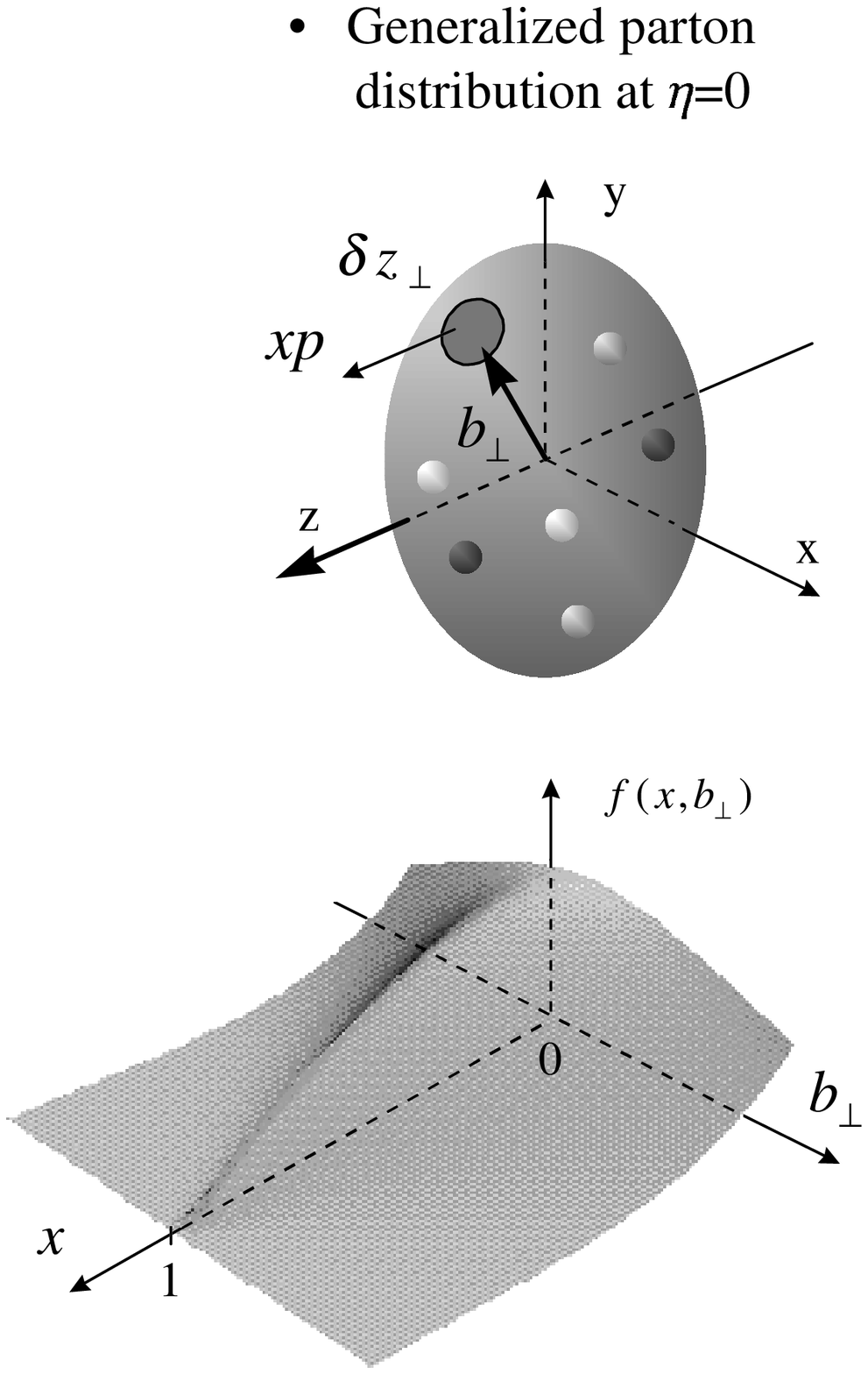}}
\end{picture}
}
\end{center}
\caption{\label{PartonContent} Probabilistic interpretation of form factors,
parton densities and generalized parton distributions at $\eta = 0$ in the
infinite momentum frame $p_z \to \infty$.}
\vspace{-0.5cm}
\end{figure}

The Breit frame is not particularly instructive for an interpretation of
high-energy scattering. Here an infinite momentum frame (IMF) is more useful,
see discussion below. In this frame, obtained by a {\sl z}-boost, the nucleon
momentum is $p_z = (p_1 + p_2)_z \to \infty$. In the IMF one builds a nucleon
state localized in the transverse plane at $\mbox{\boldmath$b$}_\perp = (x, y)$
\begin{equation}
\label{TransverseLocalNucleon}
| p_z, \mbox{\boldmath$b$}_\perp \rangle =
\sum_{\mbox{\boldmath${\scriptstyle p}$}_\perp}
\frac{
{\rm e}^{i \mbox{\boldmath${\scriptstyle p}$}_\perp
\cdot
\mbox{\boldmath${\scriptstyle b}$}_\perp}
}{
\sqrt{V_\perp}
}
{\mit\Psi} (\mbox{\boldmath$p$}_\perp)
| p_z, \mbox{\boldmath$p$}_\perp \rangle \, .
\end{equation}
Then one finds that the transverse charge distribution of the
nucleon wave packet, see Fig.\ \ref{PartonContent}, is given by the two-dimensional
Fourier transform of form factors
\begin{equation}
\langle p_z, \mbox{\boldmath$b$}_\perp = 0
| j_+ \left( \mbox{\boldmath$b$}_\perp \right) |
p_z, \mbox{\boldmath$b$}_\perp = 0 \rangle
= h_+
\int \frac{d^2 \mbox{\boldmath$\Delta$}_\perp}{(2 \pi)^2}
{\rm e}^{
- i \mbox{\boldmath${\scriptstyle \Delta}$}_\perp
\cdot
\mbox{\boldmath${\scriptstyle b}$}_\perp
}
F_1 \left( - \mbox{\boldmath$\Delta$}^2_\perp \right)
+ \dots \, .
\end{equation}
As previously one assumes a rather delocalized transverse momentum wave
function $\sum_{\mbox{\boldmath${\scriptstyle p}$}_\perp}
{\mit\Psi}^\ast (\mbox{\boldmath$p$}_\perp + \mbox{\boldmath$\Delta$}_\perp/2)
{\mit\Psi} (\mbox{\boldmath$p$}_\perp - \mbox{\boldmath$\Delta$}_\perp/2)
\approx 1$. Thus, we can interpret form factors as describing the transverse
localization of partons in a fast moving nucleon, irrespective of their
longitudinal momenta and independent on the resolution scale.

\section{Parton densities}

The deeply inelastic lepton-nucleon scattering $\ell N \to \ell' X$ probes,
via the amplitude
\begin{equation}
{\cal A}_{NX}
=
\frac{1}{Q^2} L_\mu (q) \langle p_X | j_\mu (0) | p \rangle \, ,
\end{equation}
the nucleon with the resolution $\hbar/Q \approx ( 0.2 \, {\rm fm} )/(Q \,
{\rm \, in \, GeV} )$, set by the photon virtuality $q^2 \equiv - Q^2$.
Recalling that the nucleon's size is $r_N \sim 1 \, {\rm fm}$, one concludes
that for $Q^2$ of order of a few GeV, the photon penetrates the nucleon
interior and interacts with its constituents. The cross section of the deeply
inelastic scattering is related, by the optical theorem, to the imaginary part
of the forward Compton scattering amplitude
\begin{equation}
\label{DIS-Compton}
d\sigma_{\rm DIS} \left( x_{\rm B}, Q^2 \right)
\sim
\sum_X \left| {\cal A}_{NX} \right|^2 \delta^4 (p + q - p_X)
\sim
\frac{1}{\pi} \Im{\rm m}
\ i \! \int d^4 z \, {\rm e}^{i q \cdot z}
\langle p |
T \left\{
j_\mu^\dagger (z) j_\mu (0)
\right\}
| p \rangle
\, .
\end{equation}

The very intuitive parton interpretation has its clear-cut meaning in the
IMF. A typical interaction time of partons is inversely proportional to
the energy deficit of a given fluctuation of a particle with the energy $E_0$
and three-momentum $\mbox{\boldmath$p$}_0 = (\mbox{\boldmath$p$}_{\perp 0}, x_0
p_z)$ into two partons with energies $E_{1,2}$ and three-momenta
$\mbox{\boldmath$p$}_{1,2} = (\mbox{\boldmath$p$}_{\perp 1,2}, x_{1,2} p_z)$.
It scales, for $p_z \to \infty$, as
\begin{equation}
\Delta t \sim \frac{1}{\Delta E} = \frac{1}{E_0 - E_1 - E_2}
\sim \frac{p_z}{
\mbox{\boldmath$p$}^2_{\perp 0}/x_0
-
\mbox{\boldmath$p$}^2_{\perp 1}/x_1
-
\mbox{\boldmath$p$}^2_{\perp 2}/x_2} \to \infty \, .
\end{equation}
Therefore, one can treat partons as almost free in the IMF due to the time
dilation. The virtual photon ``sees" nucleon's constituents in a frozen state
during the time of transiting the target which is, thus, describable by an
instantaneous distribution of partons. Here again the analogy with X-ray
crystallography is quite instructive: Recall that an X-ray, scattered off
atoms, reveals crystal's structure since rapid oscillations of atoms in the
lattice sites can be neglected. Atoms can be considered being at rest
during the time X-rays cross the crystal. The transverse distance probed by
the virtual photon in a Lorentz contracted hadron, is of order $\delta z_\perp
\sim 1/Q$, see Fig.\ \ref{PartonContent}. One can conclude therefore that
simultaneous scattering off an $n$-parton cascade is suppressed by an extra power
of $\left( 1/Q^2 \right)^{n - 1}$. The leading contribution to $d \sigma_{\rm DIS}$
is thus given by a handbag diagram, i.e., the photon--single-quark Compton
amplitude. The character of relevant distances in the Compton amplitude
(\ref{DIS-Compton}) is a consequence of the Bjorken limit which implies
large $Q^2$ (small distances) and energies $\nu \equiv p \cdot q$ (small
times) at fixed $x_{\rm B} \equiv Q^2/(2 \nu)$. By going to the target rest
frame one immediately finds that at large $Q^2$ the dominant contribution
comes from the light-cone distances $z^2 \approx {\cal O} \left( 1/Q^2
\right)$ between the points of absorption and emission of the virtual photon
in (\ref{DIS-Compton}) because $z_- \sim 1/(M \Bx)$, $z_+ \sim M \Bx / Q^2$.

Since the hard quark-photon subprocess occupies a very small space-time volume
but the scales involved in the formation of the nucleon are much larger, hence,
they are uncorrelated and will not interfere. The quantum mechanical incoherence
of physics at different scales results into the factorization property
of the cross section (\ref{DIS-Compton}),
\begin{equation}
\label{Factorization}
d\sigma_{\rm DIS} \left( x_{\rm B}, Q^2 \right)
\sim
\sum_q e_q^2
\int_0^1 d x \ \delta \left( x - \Bx \right) \, f_q \left( x ; Q^2 \right)
\, ,
\end{equation}
where $f_q$ is a parton distribution, --- the density of probability to
find partons of a given longitudinal momentum fraction $x$ of the parent
nucleon with transverse resolution $1/Q$,
\newpage
\begin{equation}
\langle p | \bar q (0) \gamma_+ q (z_- n) | p \rangle
=
2 p_+
\int_{0}^{1} d x
\left\{
f_q (x) {\rm e}^{- i x z_- p_+}
-
\bar f_q (x) {\rm e}^{i x z_- p_+}
\right\} \, .
\end{equation}
No information on the transverse position of partons is accessible here, Fig.\
\ref{PartonContent}.

\section{Generalized parton distributions}

\begin{figure}[t]
\begin{center}
\mbox{
\begin{picture}(0,45)(225,0)
\put(-2,-20){\insertfig{3.4}{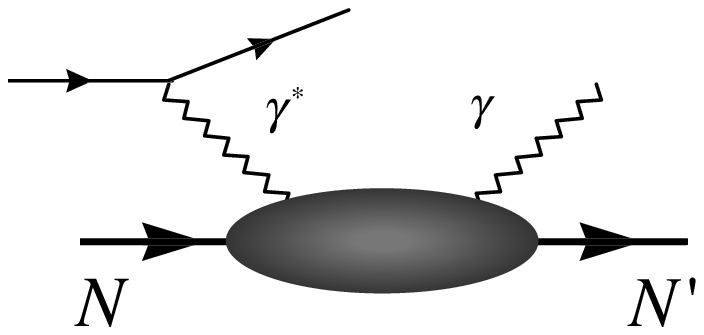}}
\put(356,-20){\insertfig{3.4}{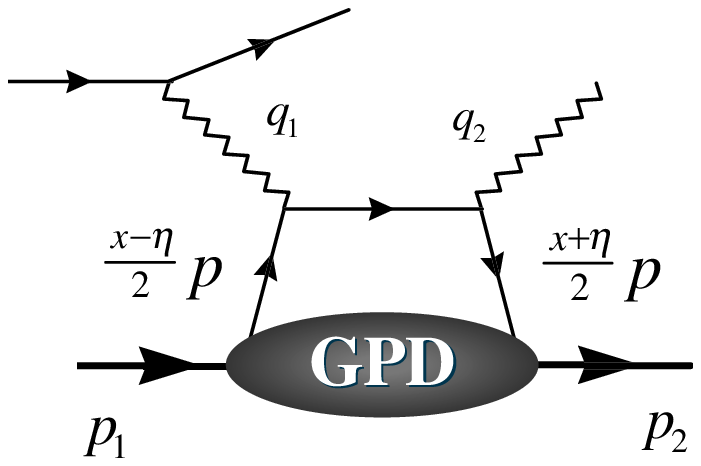}}
\put(138,-35){\insertfig{6.5}{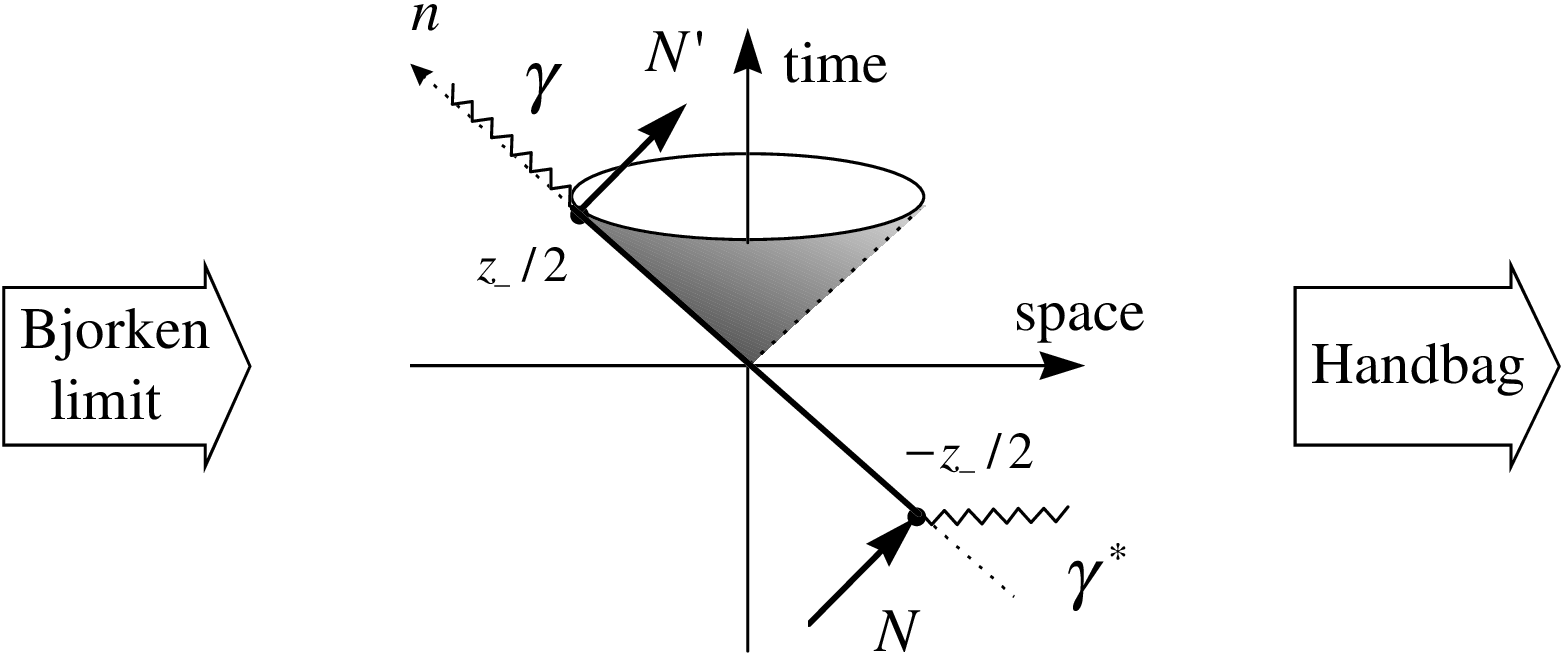}}
\end{picture}
}
\end{center}
\caption{\label{LightconeDVCS} Light-cone dominance in deeply virtual
Compton scatering.}
\vspace{-0.5cm}
\end{figure}

\begin{figure}[t]
\begin{center}
\mbox{
\begin{picture}(0,75)(100,0)
\put(0,-30){\insertfig{7}{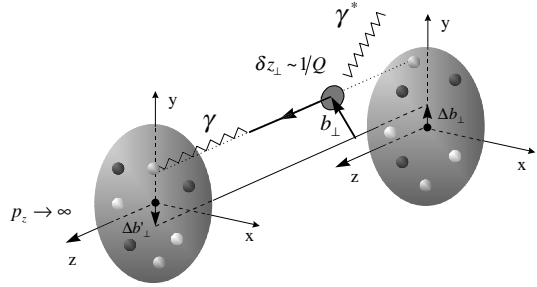}}
\end{picture}
}
\end{center}
\caption{\label{DVCSgeometry} Geometric picture of deeply virtual Compton
scattering.}
\vspace{-0.5cm}
\end{figure}

Both observables addressed in the previous two sections give only
one-dimensional slices of the nucleon since only the magnitude of
scattering amplitudes is accessed in the processes but its phase
is lost. These orthogonal spaces are probed simultaneously
in generalized parton distributions (GPDs),  which arise in the description
of deeply virtual Compton scattering (DVCS) $\ell N \to \ell' \gamma^\ast
N \to \ell' N' \gamma$ in the Bjorken limit, see Fig.\ \ref{LightconeDVCS}.
In the same spirit as in deeply inelastic scattering, the latter
consists of sending $q^2 \equiv (q_1 + q_2)^2/4 \to - \infty$ to the
deep Euclidean domain while keeping $\Delta^2 \equiv (p_2 - p_1)^2 \ll
- q^2$ small and $\xi \equiv - q^2/ p \cdot q $ fixed, $p \equiv p_1
+ p_2$. By the reasoning along the same line as in the previous section
one finds that the Compton amplitude factorizes into GPDs parametrizing
the twist-two light-ray operator matrix element
\begin{equation}
\label{defGPDs}
\langle p_2 |
\bar q (- z_- n)
\gamma_+
q (z_- n)
| p_1 \rangle
=
\int_{- 1}^{1} d x
{\rm e}^{- i x z_- p_+}
\left\{
h_+ \, H_q (x, \eta, \Delta^2)
+
e_+ \, E_q (x, \eta, \Delta^2)
\right\} \, ,
\end{equation}
and a handbag coefficient function, so that one gets from Fig.\ \ref{LightconeDVCS}
$$
{\cal A}_{\rm DVCS}
=
\varepsilon^\ast_\mu (q_2) L_\nu (q_1)
\int d^4 z \, {\rm e}^{i q \cdot z}
\langle p_2 |
T \left\{
j_\mu^\dagger (z/2) j_\nu (-z/2)
\right\}
| p_1 \rangle
\sim \sum_q e_q^2
\int_{-1}^1 dx \, \frac{F_q (x, \eta, \Delta^2)}{\xi - x - i 0} ,
$$
where $F_q = H_q, E_q$ and the contribution from a crossed diagram is
omitted. GPDs depend on the $s$-channel momentum fraction $x$, measured
with respect to the momentum $p$, and $t$-channel fraction $\eta \equiv q
\cdot \Delta/q \cdot p$, which is the longitudinal component of the
momentum transfer $\Delta \approx \eta p + \Delta_\perp$, as well as its
square $\Delta^2 \approx - \left( \mbox{\boldmath$\Delta$}_\perp^2 +
4 M_N^2 \eta^2 \right)/\left( 1 - \eta^2 \right)$. Due to the reality of
the final state photon $\eta \approx - \xi$. A geometric picture underlying
DVCS is as follows, see Fig.\ \ref{DVCSgeometry}. The electric field
of lepton's virtual fluctuation $\ell \to \ell' \gamma^\ast$ accelerates
a quark localized in the transverse area $\left( \delta z_\perp \right)^2
\sim 1/Q^2$ at the impact parameter $\mbox{\boldmath$b$}_\perp$ and carrying
a certain momentum fraction of the parent nucleon. The accelerated parton
tends to emit the energy via electromagnetic radiation and fall back into
the nucleon, see Fig.\ \ref{DVCSgeometry}. The incoming-outgoing nucleon
system is localized at the center of coordinates $\mbox{\boldmath$b$}_\perp
= 0$, however, due to  non-zero longitudinal momentum exchange in the $t$-channel
the individual transverse localizations of incoming and outgoing nucleons
are shifted in the transverse plane by amounts\footnote{Note the difference
in the definition of our impact parameter space GPDs as compared to Ref.\
\cite{Die02}. In our frame $\mbox{\boldmath$p$}_{\perp 2} = -
\mbox{\boldmath$p$}_{\perp 1} = \mbox{\boldmath$\Delta$}_\perp/2$. We define
the Fourier transform with respect to $\mbox{\boldmath$\Delta$}_\perp$ which
has its advantages that $\mbox{\boldmath$b$}_\perp$ does not depend implicitly
on $\eta$.} $\Delta \mbox{\boldmath$b$}_\perp \sim \eta (1 + \eta)
\mbox{\boldmath$b$}_\perp$ and $\Delta \mbox{\boldmath$b$}'_\perp \sim \eta
(1 - \eta) \mbox{\boldmath$b$}_\perp$, respectively \cite{Die02}.

Generally, GPDs are not probabilities rather they are the interference of
amplitudes of removing a parton with one momentum and inserting it back with
another. In the limit $\Delta = 0$ they reduce to inclusive parton densities
and acquire the probabilistic interpretation. This is exhibited in a most
straightforward way in the light-cone formalism
\cite{DieFelJakKro00BroDieHwa00}, where one easily identifies the regions
$-1<x<-\eta$ and $\eta<x<1$ with parton densities while $-\eta<x<\eta$ with
distribution amplitudes. This latter domain precludes the density
interpretation for $\eta \neq 0$.

The first moment of GPDs turns into form factors (\ref{FormFactor}). The second
moment of Eq.\ (\ref{defGPDs}) gives form factors of the quark energy-momentum
tensor, ${\mit\Theta}^q_{\mu\nu} = \bar q \gamma_{\{\mu} D_{\nu\}} q$. Since gravity
couples to matter via ${\mit\Theta}_{\mu\nu} = \frac{\delta}{\delta g_{\mu\nu}}
\int d^4 x \sqrt{{\scriptstyle - \det g_{\mu\nu} (x)}} \, {\cal L} (x)$,
these form factors are the ones of the nucleon scattering in a weak gravitational
field \cite{Ji96}
\begin{equation}
\langle p_2 | {\mit\Theta}_{\mu\nu} | p_1 \rangle
= A (\Delta^2) h_{\{ \mu} p_{\nu \}}
+ B (\Delta^2) e_{\{ \mu} p_{\nu \}}
+ C (\Delta^2) \Delta_\mu \Delta_\nu \, .
\end{equation}
Analogously to the previously discussed electromagnetic case, the combination
$A (\Delta^2) + \Delta^2/(4M_N^2) B (\Delta^2)$ arising in the ${\mit\Theta}_{00}$
component measures the mass distribution inside the nucleon \cite{BelJi02}. It is
different from the charge distribution due to presence of neutral constituents
inside hadrons not accounted in electromegnetic form factors. The gravitomagnetic
form factor $A (\Delta^2) + B (\Delta^2)$ at zero recoil encodes information
on the parton angular momentum \cite{Ji96} $\vec J = \frac{1}{V} \int d^3 \vec x
\, [ \vec x \times \vec {\mit\Theta} ] (x)$ expressed in terms of the momentum
flow operator ${\mit\Theta}_{0i} \equiv {\mit\Theta}_i$ in the nucleon and gives
its distribution when Fourier transformed to the coordinate space. These form
factor are accessible once GPDs are  measured:
\begin{equation}
\int_{-1}^1 dx \, x \, H (x, \eta, \Delta^2) = A (\Delta^2) + \eta^2 C (\Delta^2)
\, , \quad
\int_{-1}^1 dx \, x \, E (x, \eta, \Delta^2) = B (\Delta^2) - \eta^2 C (\Delta^2) \, .
\end{equation}

\begin{figure}[t]
\begin{center}
\mbox{
\begin{picture}(0,80)(220,0)
\put(0,-35){\insertfig{7}{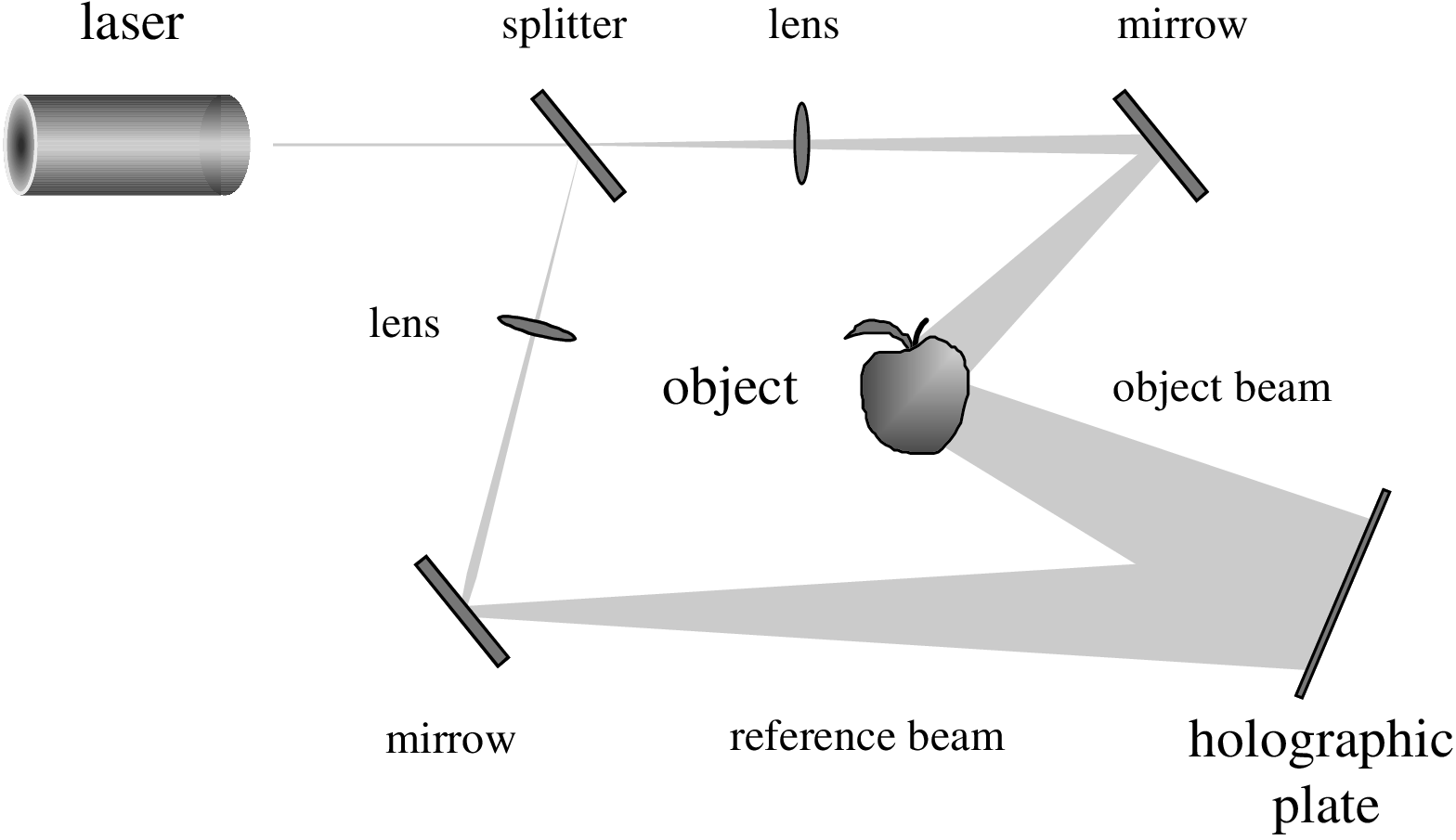}}
\put(240,-35){\insertfig{7}{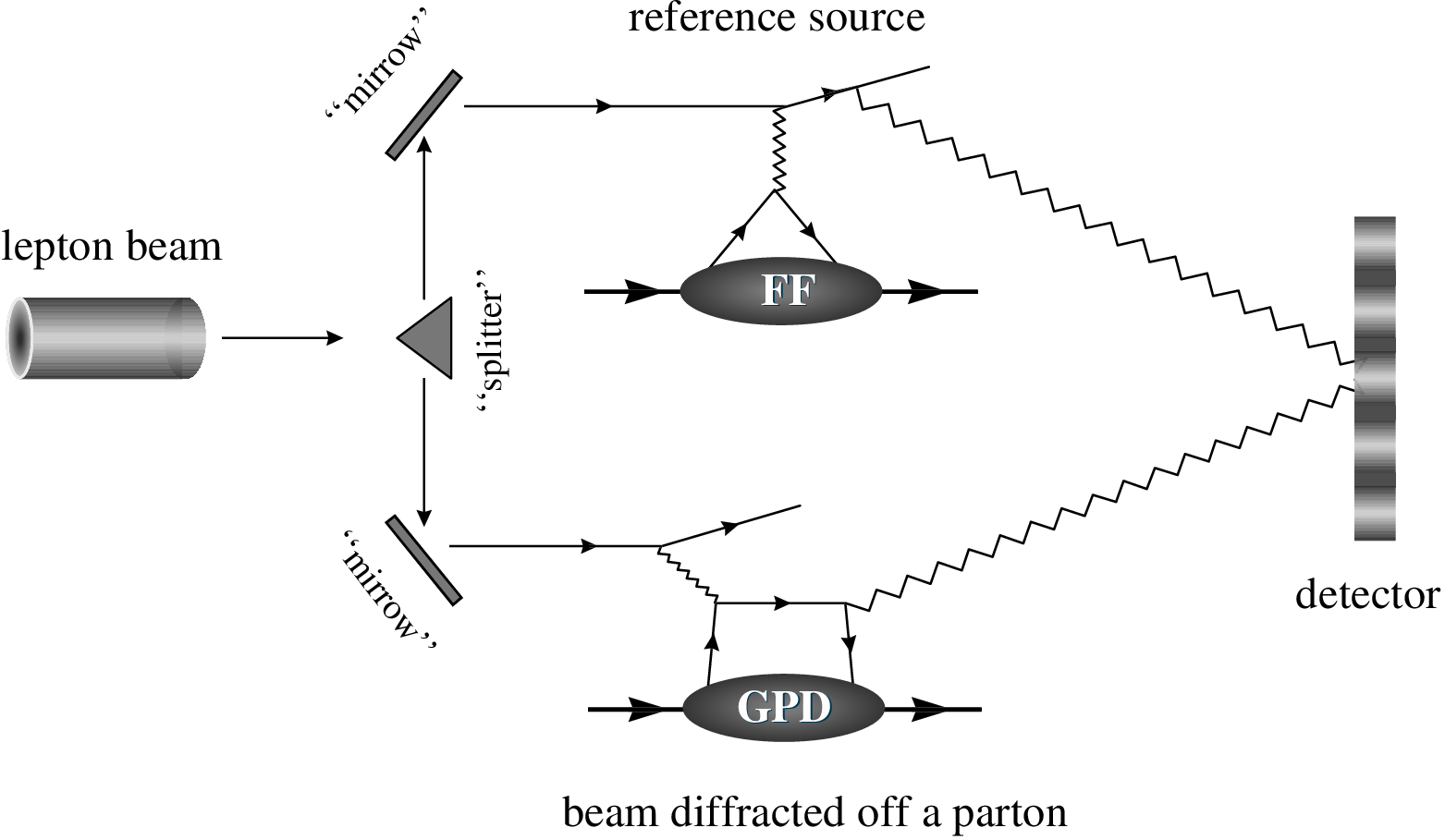}}
\end{picture}
}
\end{center}
\caption{\label{Holography} Left: Conventional setup for taking the
holographic picture. Right: Nucleon hologram with leptoproduction of
a photon: interference of the Bethe-Heitler (reference) and DVCS
(sample) amplitudes.}
\vspace{-0.5cm}
\end{figure}

GPDs regain a probabilistic interpretation once one sets $\eta = 0$ but
$\mbox{\boldmath$\Delta$}_\perp \neq 0$ \cite{Bur00,RalPir01}. When Fourier
transformed to the impact parameter space they give a very intuitive picture
of measuring partons of momentum fraction $x$ at the impact parameter
$\mbox{\boldmath$b$}_\perp$ with the resolution of order $1/Q$ set by the
photon virtuality in the localized nucleon state (\ref{TransverseLocalNucleon}),
\begin{equation}
f (x, \mbox{\boldmath$b$}_\perp)
=
\int \frac{d^2 \mbox{\boldmath$\Delta$}_\perp}{(2 \pi)^2}
{\rm e}^{
-i \mbox{\boldmath${\scriptstyle \Delta}$}_\perp
\cdot
\mbox{\boldmath${\scriptstyle b}$}_\perp
}
H \left( x, 0, - \mbox{\boldmath$\Delta$}^2_\perp \right)
\, .
\end{equation}
To visualize it, see Fig.\ \ref{PartonContent}, one can stick to the
Regge-motivated ansatz $H ( x, 0, - \mbox{\boldmath$\Delta$}^2_\perp
) \sim x^{- \alpha_R ( - \mbox{\boldmath{${\scriptstyle \Delta}$}}^2_\perp )}
(1 - x)^3$ with a linear trajectory $\alpha_R ( - \mbox{\boldmath$\Delta$}^2_\perp )
= \alpha_R (0) - \alpha'_R \mbox{\boldmath$\Delta$}^2_\perp$ where $\alpha_R (0)
\approx 0.5$ and $\alpha' \approx 1 \ \GeV^2$.

\section{Hard leptoproduction of real photon and lepton pair}

The light-cone dominance in DVCS is a consequence of the external
kinematical conditions on the process in the same way as in deeply inelastic
scattering. Therefore, one can expect precocious scaling starting as early
as at $- q^2 \sim \, 1 \, \GeV^2$. It is not the case for hard exclusive
meson production, giving access to GPDs as well, where it is the dynamical
behavior of the short-distance parton amplitude confined to a small transverse
volume near the light cone that drives the perturbative approach to the
process. Here the reliability of perturbative QCD predictions is postponed to
larger momentum transfer.

\begin{figure}[t]
\begin{center}
\mbox{
\begin{picture}(0,80)(190,0)
\put(0,-30){\insertfig{5}{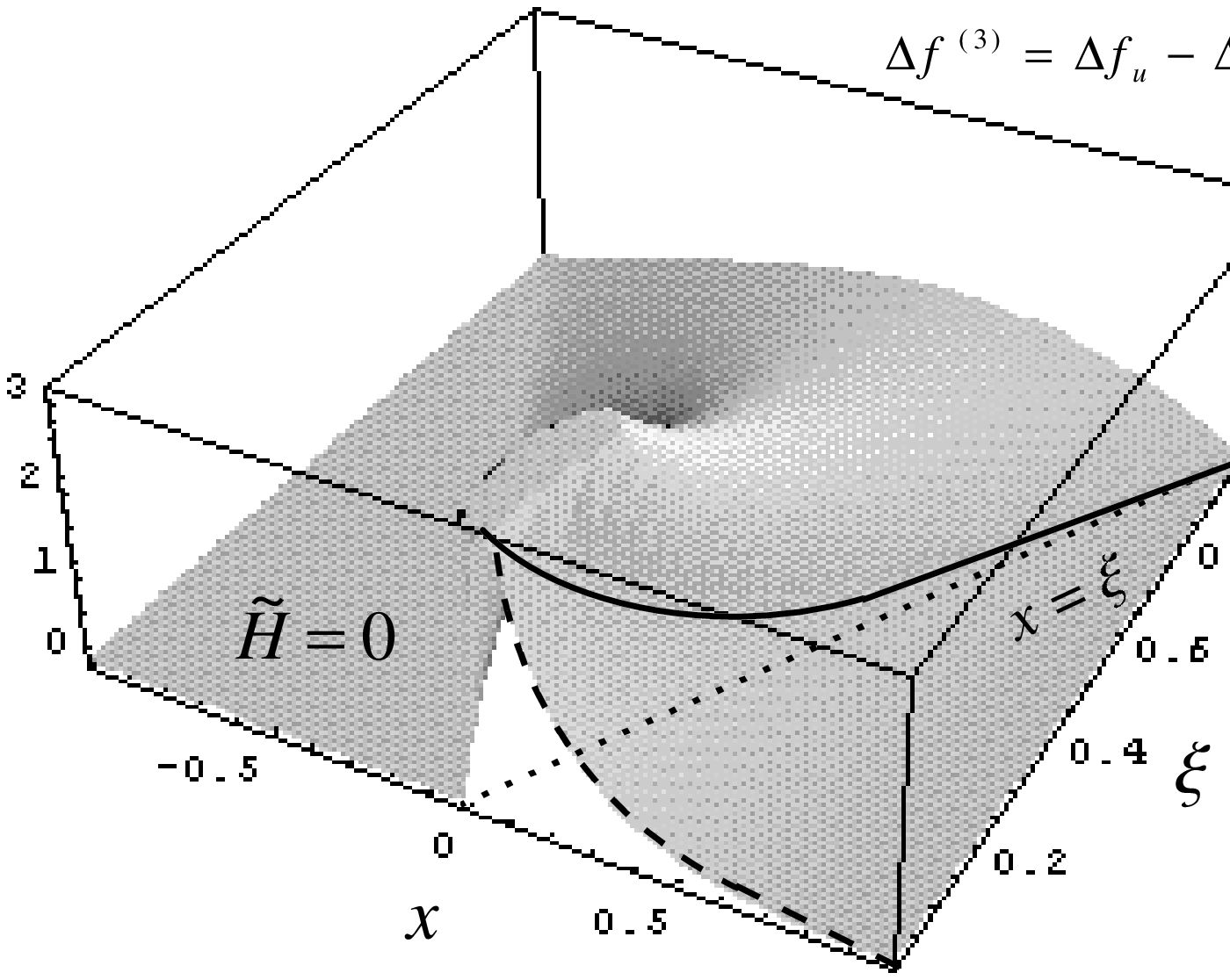}}
\put(230,-30){\insertfig{5}{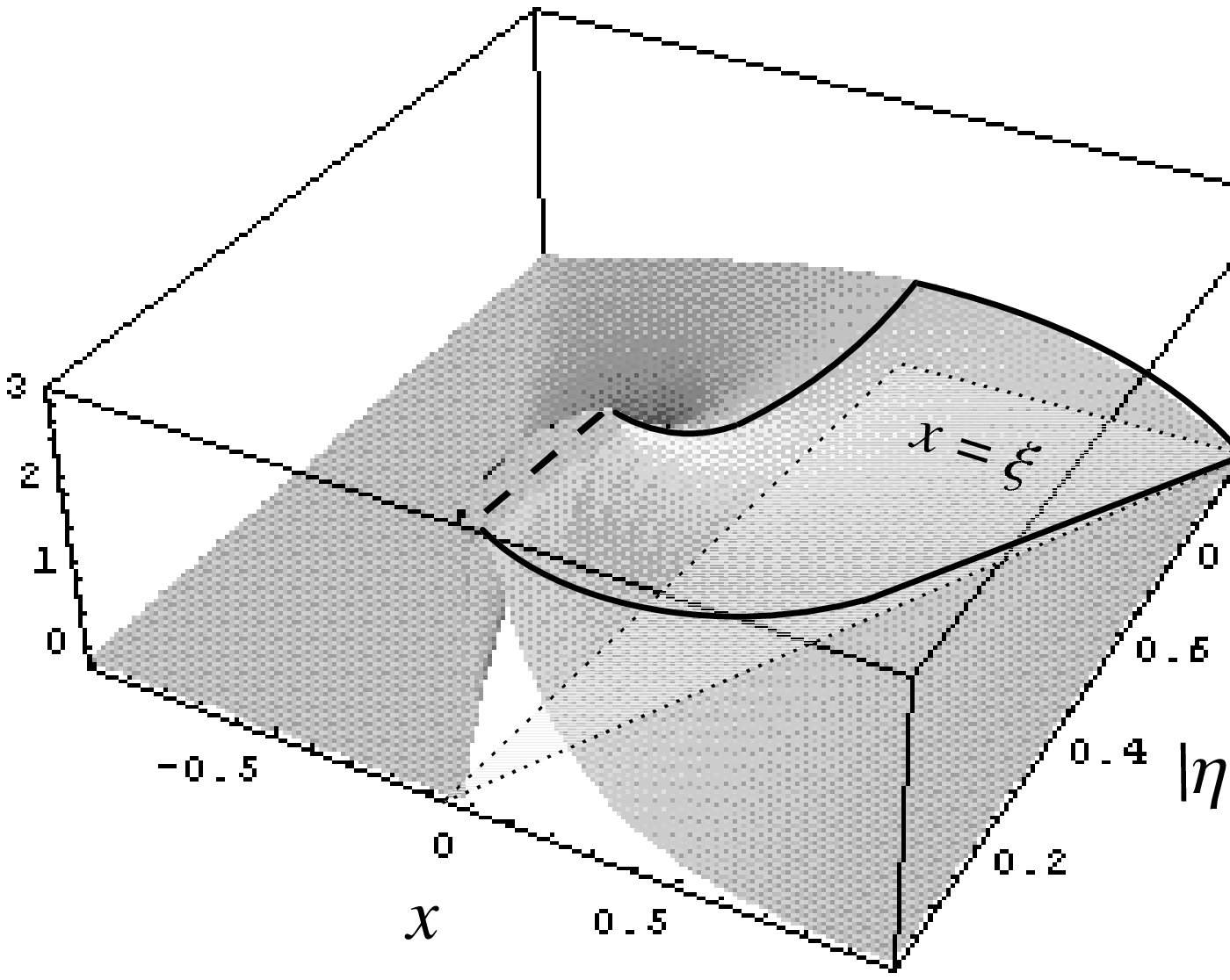}}
\end{picture}
}
\end{center}
\caption{\label{GPDfromDVCS} What is extractable from DVCS (left) and
DVCS lepton pair production (right).}
\vspace{-0.5cm}
\end{figure}

\begin{figure}[t]
\begin{center}
\mbox{
\begin{picture}(0,100)(225,0)
\put(35,80){\footnotesize (a)}
\put(0,-30){\insertfig{7.5}{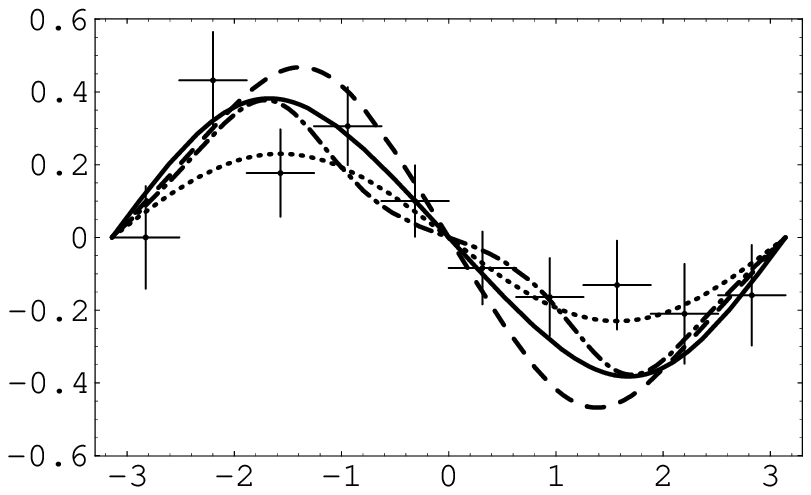}}
\put(147,87){\tiny $\Bx = 0.11$}
\put(147,77){\tiny $-\Delta^2 = 0.27\ \GeV^2$}
\put(147,67){\tiny $- q_1^2 = 2.6\ \GeV^2$}
\put(37,-10){\footnotesize HERMES DATA \protect\cite{Airetal01}}
\put(265,80){\footnotesize (b)}
\put(230,-30){\insertfig{7.5}{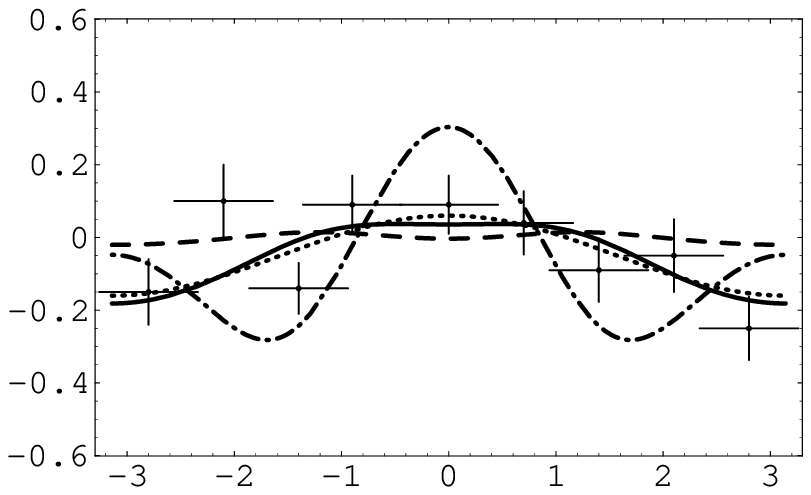}}
\put(377,87){\tiny $\Bx = 0.12$}
\put(377,77){\tiny $-\Delta^2 = 0.27\ \GeV^2$}
\put(377,67){\tiny $- q_1^2 = 2.8\ \GeV^2$}
\put(205,-35){${\scriptstyle \phi_\gamma^\prime}$}
\put(435,-35){${\scriptstyle \phi_\gamma^\prime}$}
\put(267,-10){\footnotesize PRELIMINARY HERMES DATA \protect\cite{Ell02}}
\end{picture}
}
\end{center}
\caption{\label{Fig-SSACA} Beam spin asymmetry (a) in $e^+ p \to e^+ p\gamma$
and unpolarized charge asymmetry (b) from HERMES with $E = 27.6\ \GeV$ are
predicted making use of the complete twist-three analysis for input GPDs from
Ref.\ \protect\cite{BelMueKir01}: model A without the D-term (solid) and C
with the D-term (dashed) in the Wandzura-Wilczek approximation \cite{BelMul00}
as well as the model B with the D-term (dash-dotted) and included quark-gluon
correlations. The dotted lines on the left and right panels show $0.23 \sin
\phi_\gamma^\prime$ and $-0.05 + 0.11 \cos \phi_\gamma^\prime$ HERMES fits,
respectively. Note that a toy model for quark-gluon correlations while only
slightly changing the beam asymmetry, however, strongly alter the charge
asymmetry.}
\vspace{-0.7cm}
\end{figure}

Although GPDs carry information on both longitudinal and transverse degrees
of freedom, their three-dimensional experimental exploration requires
a complete determination of the DVCS amplitude, i.e., its magnitude and phase.
One way to measure the phase at a given spot is known as holography, for
visible light. This technology allows to make three-dimensional photographs
of objects, see Fig.\ \ref{Holography}: The laser beam splits into two rays.
One of them serves as a reference source and the other reflects from the
object's surface. The reflected beam, which was in phase with the reference
beam before hitting the ``target", interferes with the reference beam and
forms fringes on the plate with varying intensity depending on the phase
difference of both. (Unfortunately, the same method cannot be used for
X-ray holography of crystals and scattering experiments due to the absence
of practical ``splitters".) For the exclusive leptoproduction of a photon,
however, there are two contributions to the amplitude: the DVCS one
${\cal A}_{\rm DVCS}$,  we are interested in,  and ${\cal A}_{\rm BH}$ from
the `contaminating' Bethe-Heitler (BH) process, in which the real photon
spills off the scattered lepton rather than the quark, see Fig.\ \ref{Holography}.
The BH amplitude is completely known since the only long-distance input turns
out to be nucleon form factors measured elsewhere. The relative phase of
the amplitudes can be measured by the interference of DVCS and BH amplitudes
in the cross section $d\sigma_{\ell N \to \ell' N' \gamma} \sim \left|
{\cal A}_{\rm DVCS} + {\cal A}_{\rm BH} \right|^2$ and, thus, the nucleon
hologram can be taken. The most straightforward extraction of the interference term is
achieved by making use of the opposite lepton charge conjugation properties
of DVCS and BH amplitudes. The former is odd while the latter is even under
change of the lepton charge. The unpolarized beam charge asymmetry gives
\begin{eqnarray*}
d\sigma_{\ell N \to \ell' N' \gamma} (+ e_\ell)
-
d\sigma_{\ell N \to \ell' N' \gamma} (- e_\ell)
=
\left( {\cal A}_{\rm DVCS} + {\cal A}^\ast_{\rm DVCS} \right)
{\cal A}_{\rm BH}
\sim
\Re{\rm e} \int \frac{F_q (x, \xi, \Delta^2)}{\xi - x - i 0}
\cos \phi^\prime_\gamma
\hspace{-1cm}
\end{eqnarray*}
and measures the real part of the DVCS amplitude modulated by the harmonics of
the azimuthal angle between the lepton and photon scattering planes $\phi'_\gamma$
\cite{BelMueKir01,DieGouPirRal97BelMueNieSch00}. If on top of the charge asymmetry
one further forms either beam or target polarization differences, this procedure
would allow to cleanly extract the imaginary part of the DVCS amplitude where
GPDs enter in diverse combinations. These rather involved measurements have not
yet been done. Luckily, since the ratio of BH to DVCS amplitude scales like
$[\Delta^2/q_1^2 (1 - y)]^{1/2}/y$, for large $y$ or small $- \Delta^2$, it is safe
to neglect $|{\cal A}_{\rm DVCS}|^2$ as compared to other terms. Thus, in such
kinematical settings one has access to the interference in single spin asymmetries,
\begin{eqnarray*}
d\sigma_{\ell N \to \ell' N' \gamma} (+ \lambda_\ell)
-
d\sigma_{\ell N \to \ell' N' \gamma} (- \lambda_\ell)
\approx
\left( {\cal A}_{\rm DVCS} - {\cal A}^\ast_{\rm DVCS} \right)
{\cal A}_{\rm BH}
\sim
\Im{\rm m} \int \frac{F_q (x, \xi, \Delta^2)}{\xi - x - i 0}
\sin \phi^\prime_\gamma
\hspace{-1cm}
\end{eqnarray*}
which measure GPDs directly on the line $x = \xi$ as shown in Fig.\ \ref{GPDfromDVCS}.
Experimental measurements of these asymmetries were done by HERMES
\cite{Airetal01,Ell02} and CLAS \cite{Steetal01} collabarations. The comparison
to current GPD models is demonstrated in Fig.\ \ref{Fig-SSACA}.

In order to go off the diagonal $x = \xi$ one has to relax the reality
constraint on the outgoing $\gamma$-quantum, i.e., it has to be virtual and
fragment into a lepton pair $\bar L L$ with invariant mass $q_2^2 > 0$.
Thus, one has to study the process $\ell N \to \ell' \bar L L N'$. In these
circumstances, the skewedness parameter $\eta$ independently varies  for fixed
Bjorken variable since $\xi \approx - \eta ( |q_1^2| - q_2^2 )/( |q_1^2| + q_2^2 )$,
and one is able to scan the three-dimensional shape of GPDs, see Fig.
\ref{GPDfromDVCS}. Unfortunately, the cross section for DVCS lepton
pair production is suppressed by $\alpha_{\rm em}^2$ as compared to DVCS and
also suffers from resonance backgrounds, see, e.g., \cite{BerDiePir01}.

Finally, perturbative next-to-leading (NLO) and higher-twist effects are
shortly discussed. Estimates of the former are, in general, model dependent.
NLO contributions to the hard-scattering amplitude \cite{BelMue97a} of a
given quark species are rather moderate, i.e., of the relative size of $20\%$,
however, the net result in the DVCS amplitude can be accidentally large
\cite{BelMueKir01,FreMcD01b}. This can be caused by a partial cancellation
that occurs in tree amplitudes. Evolution effects \cite{Mue94} in the flavor
non-singlet sector are rather small. In the case of gluonic GPD models we
observed rather large NLO corrections to the DVCS amplitude for the naive scale
setting $\mu_F^2 = -q_1^2$ \cite{BelMueKir01}. For such models one also
has rather strong evolution effects, which severely affect LO analysis.
However, one can tune the factorization scale $\mu_F$ so that to get rid
of these effects. The renormalon-motivated twist-four \cite{BelSch98} and
target mass corrections \cite{BelMue01} await their quantitative exploration.



\end{document}